\documentclass[conference]{IEEEtran}
\usepackage{cite}
\usepackage{amsmath,amssymb,amsfonts}
\usepackage{algorithmic}
\usepackage{graphicx}
\usepackage{textcomp}
\usepackage{xcolor}
\usepackage{fancyhdr}
\usepackage[hyphens]{url}
\usepackage{authblk}
\usepackage{subcaption}

\usepackage{soul}

\def\BibTeX{{\rm B\kern-.05em{\sc i\kern-.025em b}\kern-.08em
    T\kern-.1667em\lower.7ex\hbox{E}\kern-.125emX}}

\fancypagestyle{firstpage}{
  \fancyhf{}
  
  \fancyfoot[C]{\thepage}
}

\title{GPU Domain Specialization via Composable On-Package Architecture} 
\author{Yaosheng Fu, Evgeny Bolotin, Niladrish Chatterjee, David Nellans, Stephen W. Keckler}
\affil{NVIDIA}

\begin{document}
\maketitle
\thispagestyle{firstpage}
\pagestyle{plain}


\begin{abstract}
As GPUs scale their low precision matrix math throughput to
boost deep learning (DL) performance, they upset the balance between
math throughput and memory system capabilities. We demonstrate that
converged GPU design trying to address diverging architectural
requirements between FP32 (or larger) based HPC and FP16 (or smaller)
based DL workloads results in sub-optimal configuration for either of
the application domains. We argue that a \textbf{C}omposable
\textbf{O}n-\textbf{PA}ckage \textbf{GPU} (COPA-GPU) architecture to
provide domain-specialized GPU products is the most practical solution
to these diverging requirements. A COPA-GPU leverages
multi-chip-module disaggregation to support maximal design reuse,
along with memory system specialization per application domain. We
show how a COPA-GPU enables DL-specialized products by modular
augmentation of the baseline GPU architecture with up to 4$\times$
higher off-die bandwidth, 32$\times$ larger on-package cache,
2.3$\times$ higher DRAM bandwidth and capacity, while conveniently
supporting scaled-down HPC-oriented designs. This work explores the
microarchitectural design necessary to enable composable GPUs and
evaluates the benefits composability can provide to HPC, DL training,
and DL inference. We show that when compared to a converged GPU design,
a DL-optimized COPA-GPU featuring a combination of 16$\times$
larger cache capacity and 1.6$\times$ higher DRAM bandwidth scales
per-GPU training and inference performance by 31\% and 35\%
respectively and reduces the number of GPU instances by 50\% in
scale-out training scenarios.
\end{abstract}

\section{Introduction}
\label{sec:intro}
Deep learning (DL) has revolutionized computer vision, natural
language processing, speech recognition, and recommendation
systems~\cite{resnet50_cvpr_2016,bert_2018,ds2_icml_2016,ncf_www_2017}
and reshaped the automotive, robotics, e-commerce, and healthcare
industries~\cite{av_2017,unet_2019,id_nips_2014,robot_icra_2017}\@.
GPUs have become the de facto platform of choice for DL applications
due to their relative ease of programming, rich set of optimized
software libraries, and good balance between compute and off-chip
memory bandwidth~\cite{ampere}\@. GPU acceleration can reduce the
training time of DL applications from months to
minutes~\cite{mlperf_training_2019}\@. However, GPU architectures must
keep evolving to satisfy the unprecedented expansion in DL computing
demand, fueled by both the growing complexity of model and dataset
sizes~\cite{nlpscaling_2020}\@.

GPU vendors have long enjoyed the benefits of shrinking process
technologies that increased transistor density, while also
manufacturing ever larger
dies~\cite{moorecramming,kepler,pascal,volta,ampere}\@. Thanks to
technology scaling, architects have had ample resources to add
domain-specialized hardware to computer graphics-oriented GPUs,
resulting in today's converged GPU design that serves multiple
application domains. To cater to High Performance Computing (HPC),
GPUs added costly high precision arithmetic (FP64) units, advanced
error detection and correction hardware, high bandwidth memories
(HBM), and fast I/O such as NVIDIA's NVLINK~\cite{volta, ampere,
amd_mi100, nvlink}\@. Similarly, specialized hardware such as NVIDIA
Tensor Cores and AMD Matrix Cores were added for DL, while RT Cores
were added to accelerate next generation ray tracing
capabilities~\cite{volta, ampere, turing, amd_mi100}\@.

\begin{figure}[t]
  \center
  \includegraphics[width=1\linewidth]{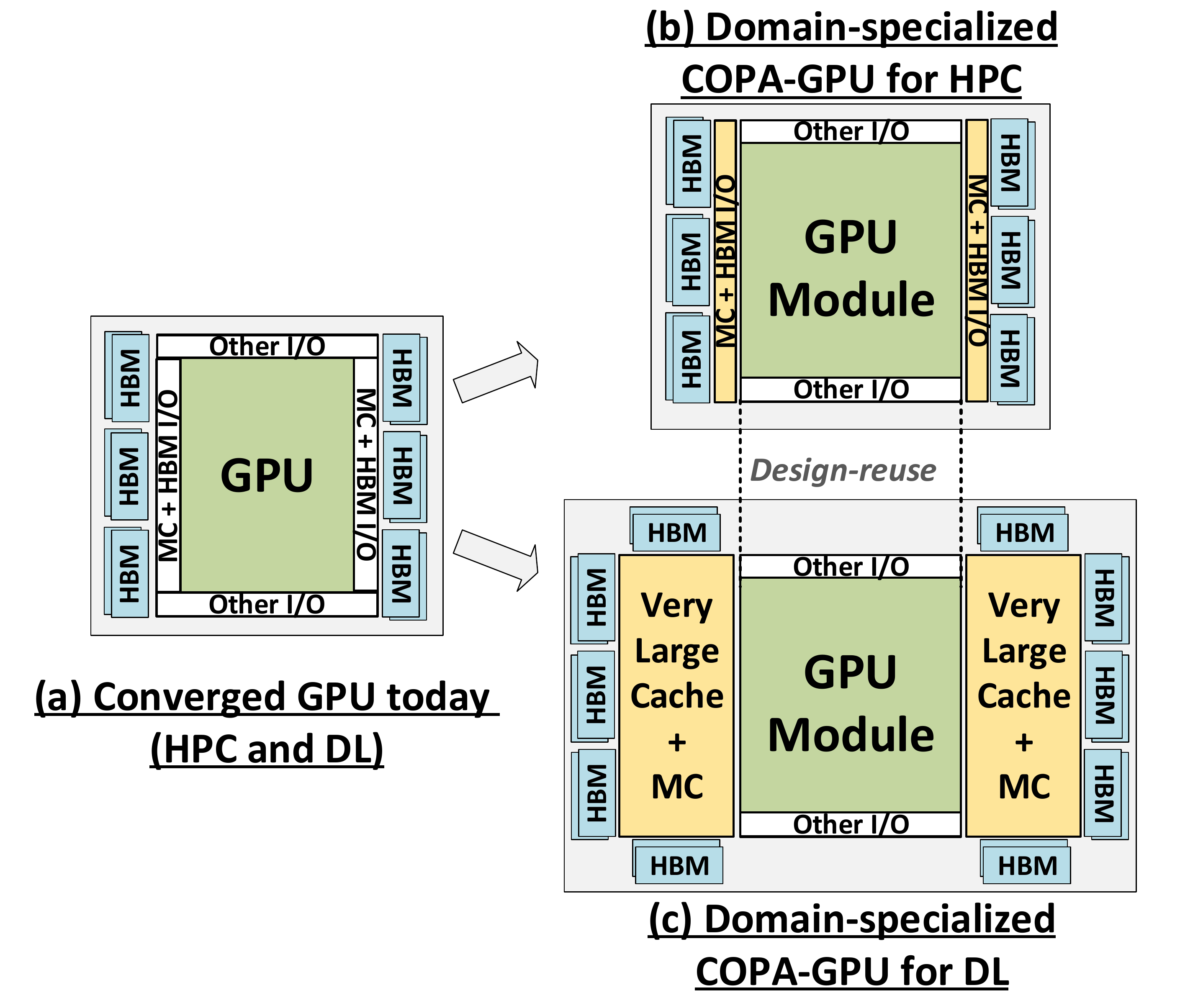}
\caption{(a) A converged monolithic GPU for HPC and DL. (b) An HPC-specialized COPA-GPU with similar capabilities to converged GPU.
(c) A DL-specialized COPA-GPU featuring a large capacity, high-bandwidth
on-package cache, and scaled DRAM resources.}
\label{fig:copa_gpu_hl}
  \vspace{-0.2in}
\end{figure}

The recent dramatic improvement in low precision math throughput on
GPUs is driven by research work showing the feasibility of high accuracy DL
training and inference while utilizing lower precision
arithmetic~\cite{mixedtraining_iclr_2018,bnn_nips_2016,tnn_2016}\@.
For example, NVIDIA's V100 GPU introduced DL Tensor Cores improving
FP16 throughput by 6$\times$ in one GPU generation~\cite{volta}\@. NVIDIA's
Turing GPU family went even further, integrating INT4/8 support into
the Tensor Cores~\cite{turing}\@. Most recently, NVIDIA's A100 GPU
increased the FP16 throughput by an additional 2.5$\times$ over
V100~\cite{ampere}\@.

This work demonstrates that converged GPU designs are on track to become
``over-designed`` in terms of memory bandwidth to compute throughput
for FP32/FP64-based applications while also severely ``under-designed`` due
to the rapid scaling in low precision math capabilities utilized
by next generation DL-networks.
Moreover, the slowdown in transistor scaling
along with reticle limitations in silicon lithography are restricting
transistor and die area growth, making it ever harder to deliver
competitive performance in multiple application domains via a single
converged GPU design. Future off-chip bandwidth scaling is also at
risk as the die edge scales sub-linearly with increasing die size,
limiting the off-chip DRAM and I/O bandwidth available to individual
dies~\cite{MCM-GPU}\@. 

The growing importance of DL domain specific acceleration in the datacenter
is demonstrated by a proliferation of accelerators targeted
exclusively at DL
algorithms~\cite{TPU_isca_2017,graphcore_2019,cerebras_2019,groq_2020}\@.
We believe that the combination of technology trends and an
ultra-competitive DL landscape has created an inflection point at which
GPU manufacturers must embrace more domain-specialized GPU designs to
continue providing competitive performance, while also maximizing
design reuse and minimizing non-recurring engineering (NRE) costs.


In this work we examine historical trends and project future
ratios, that show a strong divergence between high and low precision
math throughput on GPUs\@. We show that this gap makes it difficult to
design a memory system that efficiently supports both HPC and DL use
cases. Thus, we propose a \textbf{C}omposable
\textbf{O}n-\textbf{PA}ckage \textbf{GPU} (COPA-GPU) architecture as a
practical solution for building a new class of domain-optimized GPUs.
Leveraging multi-chip module (MCM) integration~\cite{MCM-GPU,
AMDZEPPELIN, COWOS,NUMA-GPU, AMD-chiplets-isscc20,Kannan15} along with
emerging circuit technology
innovations~\cite{onpackage_memsys_2015,COWOS,ARMCOWOS}, we propose to
replace the single converged GPU that serves both HPC and DL domains
with composable, semi-specialized designs as shown in
Figures~\ref{fig:copa_gpu_hl}(b) and (c).

A COPA-GPU that is specialized for HPC (similar to
Figure~\ref{fig:copa_gpu_hl}(b)) uses the baseline GPU module (GPM)
and a memory system modules providing similar or scaled down capabilities as in the converged GPU today. 
Augmenting the memory system modules with a very
large on-package cache, accessible at high on-package bandwidth and
additional off-chip DRAM resources
(Figure~\ref{fig:copa_gpu_hl}(c)) results in a substantially different
COPA-GPU design point that is well suited for bandwidth-hungry DL workloads.
This level of specialization is not free, as it requires both
intelligent architectural disaggregation of the GPU memory system and
employing emerging circuit and packaging techniques. This work
aims to shed light on the architectural choices available to designers when
building and specializing composable GPU designs.

We make the following contributions:

\begin{itemize}
  
  \item We examine historical GPU trends and perform a detailed performance
  analysis of diverging HPC and DL benchmarks. We show that due to the
  converged nature of GPUs and future DL scaling requirements, the GPU's memory bandwidth will become the primary performance bottleneck
  for GPU based DL training and inference, while being under-utilized for most HPC applications.
  
  \item We propose the development of domain-specialized composable
  GPU architectures. COPA-GPUs provide high levels of GPU design reuse
  across the HPC and DL domains, while enabling specifically optimized
  products for each domain. We describe both the architectural
  modifications, as well as the landscape of technologies, needed to enable
  COPA-GPUs.


\item We evaluate the performance potential of COPA-GPU in the context
of DL training and inference and show that very large cache
capacity can dramatically improve DL-inference, but both
cache and DRAM improvements (available only through COPA designs)
are necessary to significantly improve DL-training.

\item Finally, we propose a specific COPA-GPU design boosting
per-GPU training and inference performance by up to 31\% and 35\%
respectively, while also significantly reducing the cost of scale-out
GPU training in datacenters.

\end{itemize}

\vspace{-0.05in}
\section{Motivation and Background}
\label{sec:motivation}
The commercial importance of deep learning is undeniable and numerous
companies are now designing targeted DL training and inference
accelerators, such as Google's TPU~\cite{TPU_isca_2017, TPU_2020},
Graphcore's IPU~\cite{graphcore_2020}, Cerebras'
WSE~\cite{cerebras_2019}, and Groq's TSP~\cite{groq_2020}\@. These
application-specific architectures need not maintain a legacy of
high performance in other domains, allowing them to make heavily specialized
architectural choices compared to GPU architectures that currently serve
multiple domains.  To better understand the divergence of GPU and DL
accelerator design trends we examined recent GPU scaling trends,
projected a plausible future GPU configuration, and now compare it to
several dedicated DL accelerators.

\begin{table}[t]
    \centering
    \caption{GPU and DL accelerator compute and memory system trends,
     including a forward-looking GPU projection (GPU-N). We report the
     attributes of Cerebras as 1/84 tiles in a wafer to provide
     area-similar comparisons.}
    \label{tab:gpu_trend}
    \scriptsize
    \renewcommand\tabcolsep{4pt}
    \begin{tabular}{|l|c|c|c|c|c|}
    \hline
    \multicolumn{1}{|c|}{GPU}           & FP32     & FP16     & L2      &-    & DRAM BW   \\ 
    \multicolumn{1}{|c|}{Architecture}  & [TFLOPS] & [TFLOPS] & [MB]    &    & [GB/s] \\
    \hline
    NVIDIA P100        & 11      & 21        & 4            &-   & 732 \\ \hline
    NVIDIA V100         & 16      & 125       & 6            &-  & 900  \\ \hline
    NVIDIA A100        & 20      & 312       & 40           &-  & 1,555 \\\hline
    GPU-N       & 24      & 779       & 60           &-  & 2,687 \\ \hline
    \hline
                    &      &      & On-chip     & On-chip        &    \\ 
    \multicolumn{1}{|c|}{DL} & FP32     & FP16     & memory     & memory BW        &   DRAM BW  \\ 
    \multicolumn{1}{|c|}{Accelerator}   & [TFLOPS] & [TFLOPS] & [MB]         &  [TB/s]        & [GB/s]  \\ 
    \hline
    Graphcore IPU   & 31      & 125     & 304      & 45                & N/A \\  \hline
    Graphcore IPUv2 & -      & 250     & 900      & 180                & N/A \\  \hline
    Cerebras WSE    & - & - & 214      & 107               & N/A   \\  \hline
    Groq TSP        & - & 250     & 220      & 80                & N/A \\ \hline
    Google TPUv2    & 5      & 46       & 37       & -           & 700 \\ \hline
    Google TPUv3    & 14     & 123      & 37       & -           & 900 \\ 
    \hline
    \end{tabular}
\end{table}


\vspace{-0.05in}
\subsection{DL Architecture Trends and Projections}
\label{sec:sec:projections}

Table~\ref{tab:gpu_trend} summarizes several key compute and memory
system characteristics across the three most recent generations of
NVIDIA Tesla GPUs targeting both HPC and DL domains. We then
forward-project the hardware capabilities of a hypothetical next-GPU
configuration (GPU-N) using evolutionary scaling.  We calculate the
compute and memory bandwidth of GPU-N by linearly extrapolating these
parameters from V100 to A100~\cite{volta, ampere}\@.  We calculate the
GPU-N L2 capacity using the lower scaling rate of 1.5$\times$ from
P100~\cite{pascal} to V100 (instead of 6.7$\times$ from V100 to A100)
since GPU-N is highly unlikely to be able to fit hundreds of MB of
SRAM on a single GPU die. Note that we are not trying to accurately project the
exact configuration of a future GPU, but rather demonstrate
the effect of expected scaling rates on compute and memory bandwidths.
Table~\ref{tab:gpu_trend} makes it clear that while the memory
bandwidth to FP32 throughput ratios have generally increased across
GPU generations (from 67$\times$ in P100 to 112$\times$ in GPU-N), the memory bandwidth to FP16 ratio has
diminished to only 3.4$\times$ in GPU-N. 

In addition to several GPU generations, Table~\ref{tab:gpu_trend}
provides attributes of several DL accelerators. The Graphcore,
Cerebras, and Groq products all feature very large on-chip SRAMs as
main memory, while Google TPUs use off-chip DRAM (HBM)\@.  In contrast
to GPUs and TPUs, DL accelerators with DRAM-less memory system can
provide substantially higher memory bandwidth to FP16 ratios reaching
720$\times$ and 320$\times$ in IPUv2 and TSP respectively, albeit
providing much lower total memory capacity.

\begin{figure}[t]
 \center
 \includegraphics[width=1\linewidth]{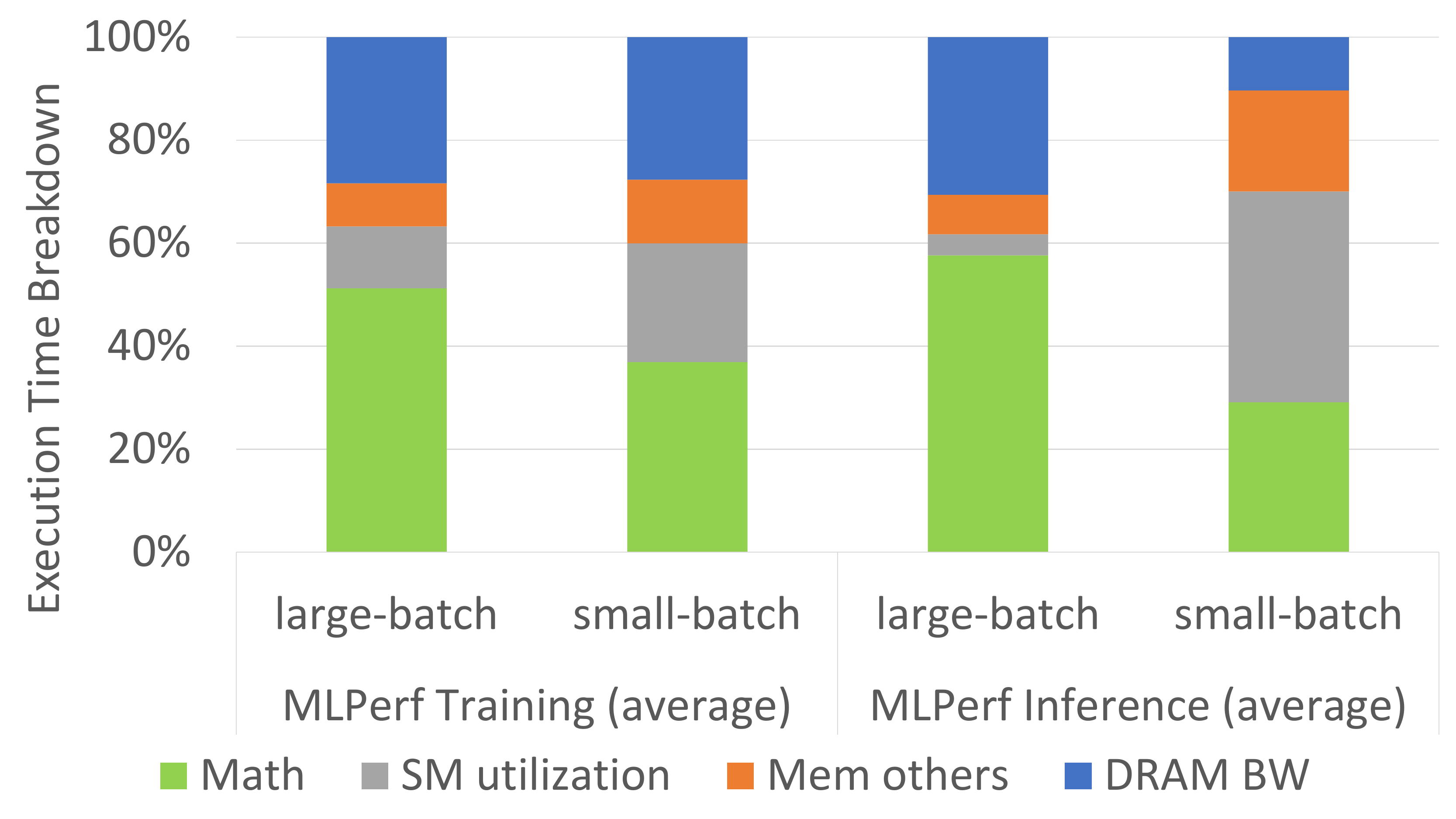}
 \caption{GPU-N performance bottleneck analysis using the MLPerf DL training and inference suites, for both large-batch and small-batch settings.}
 \label{fig:dl_bottleneck}
 \vspace{-0.2in}
\end{figure}

\vspace{-0.05in}
\subsection{Diverging DRAM Bandwidth Requirements}
\label{sec:sec:dram_bw_req}

To understand the diverging DRAM bandwidth requirements of future GPUs,
Figure~\ref{fig:dl_bottleneck} presents a simulation performance bottleneck
analysis of the DL workloads from MLPerf suite (details described
later in Section~\ref{sec:sec:methodology}) for both small- and
large-batch scenarios on GPU-N. To correctly identify the source of
performance inefficiency, we breakdown the total execution time to
individual sets of hardware components, where each bar segment
represents the performance overhead introduced by that component
during execution. For example, the blue ''DRAM BW'' bars show the
performance overhead attributed to non-ideal DRAM bandwidth, when
compared to infinite DRAM bandwidth. Similarly, the orange bars
represent the performance penalty caused by all other components in
the memory subsystem being non-ideal in terms of bandwidth and
latency. The gray bars represent the performance penalty caused by
dynamic SM under-utilization (idle SMs) versus an ideal GPU with 100\%
SM utilization, reflecting inefficiencies such as imperfect work
scheduling or lack of sufficient parallelism (sometimes only in phases) within the
workload itself. Finally, the green bars show the relative execution
time spent in Math units, which should be 100\% utilized in the ideal case.

Figure~\ref{fig:dl_bottleneck} shows that indeed DRAM bandwidth is the
primary performance limiter (excluding Math) for DL training in GPU-N,
contributing 28\% of the total execution time on average across
large- and small-batch cases. DRAM bandwidth is also the main
performance bottleneck for large-batch DL inference contributing
30\% of total execution time.  However, at small batch size SM
under-utilization accounts for 41\% of total execution time and
serves as the primary performance bottleneck rather than DRAM
bandwidth. This is because MLPerf small-batch inference does not
expose enough parallelism to fill an entire GPU that was designed for the
datacenter. Moreover, due to small-batch inference's relatively small
memory footprints, the majority of each workload's data can be
buffered on-chip to avoid DRAM bandwidth to become the bottleneck.  

\begin{figure}[t]
    \center
    \includegraphics[width=1\linewidth, trim= 0 0 0 0,clip]{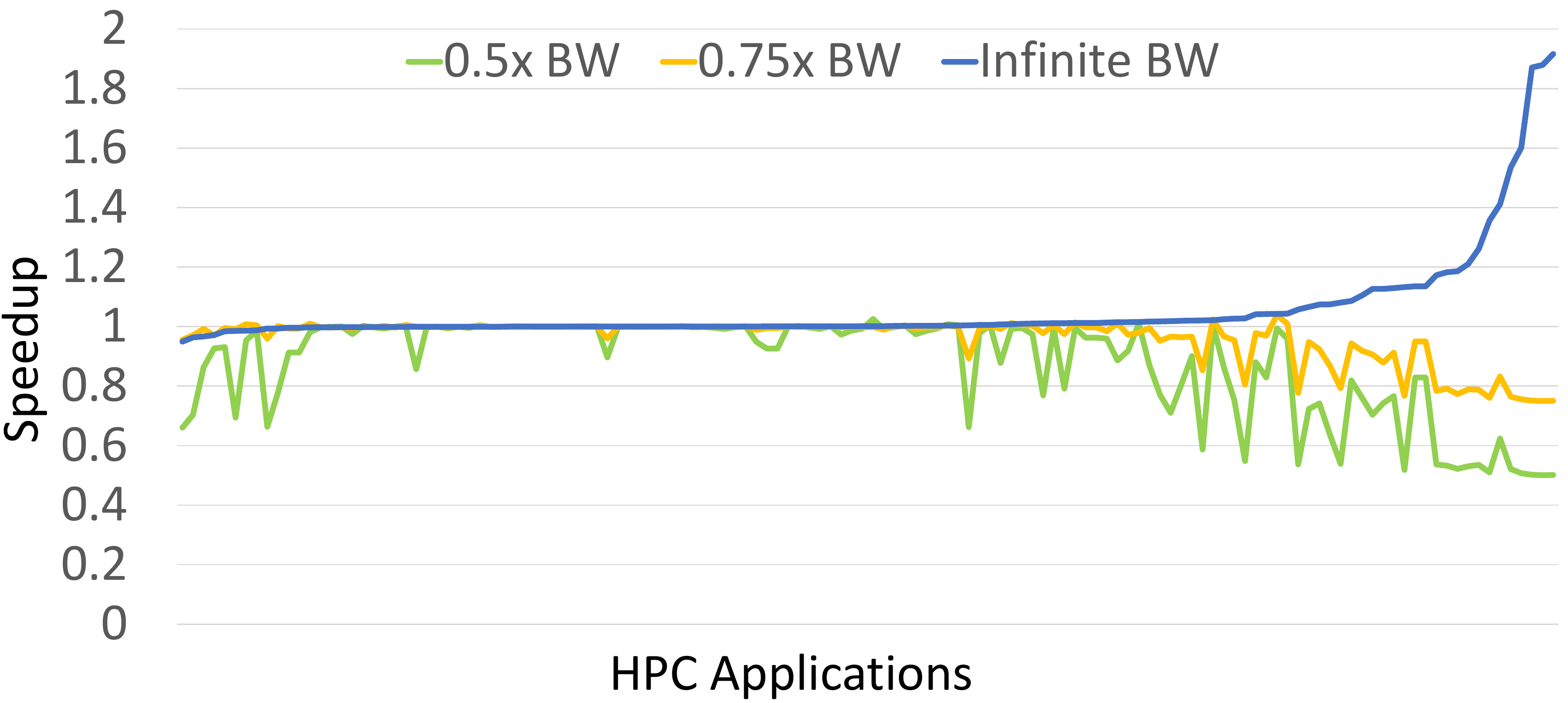}
    \vspace{-0.2in}
    \caption{Performance speedup of various DRAM bandwidth for HPC applications on a GPU-N configuration.}
    \label{fig:hpc_drambw}
    \vspace{-0.1in}
\end{figure}

We also simulate the converged GPU-N configuration across varying DRAM
bandwidth settings using 130 HPC benchmarks from differing sources
including the CORAL~\cite{CORAL_2014} and CORAL-2~\cite{CORAL2_2017}
benchmarks, Amber18 benchmarks~\cite{Amber18}, FUN3D~\cite{FUN3D},
SPECFEM3D Cartesian~\cite{SPECFEM3D}, GROMACS~\cite{GROMACS},
Laghos~\cite{laghos}, and RELION~\cite{RELION}. As shown in
Figure~\ref{fig:hpc_drambw}, in stark contrast to DL applications,
most HPC applications are quite insensitive to changes in DRAM bandwidth. 
When DRAM bandwidth is increased to infinite, a
geometric mean speedup of only 5\% is achieved. When DRAM bandwidth is
decreased, 0.75$\times$BW and 0.5$\times$BW result in 4\% and 14\%
slowdown, respectively. This implies that future increases in DRAM bandwidth will go
largely underutilized by most HPC applications if converged
GPU designs targeting both DL and HPC domains remain the de facto standard.

\begin{figure}[t]
  \center
  \includegraphics[width=1\linewidth]{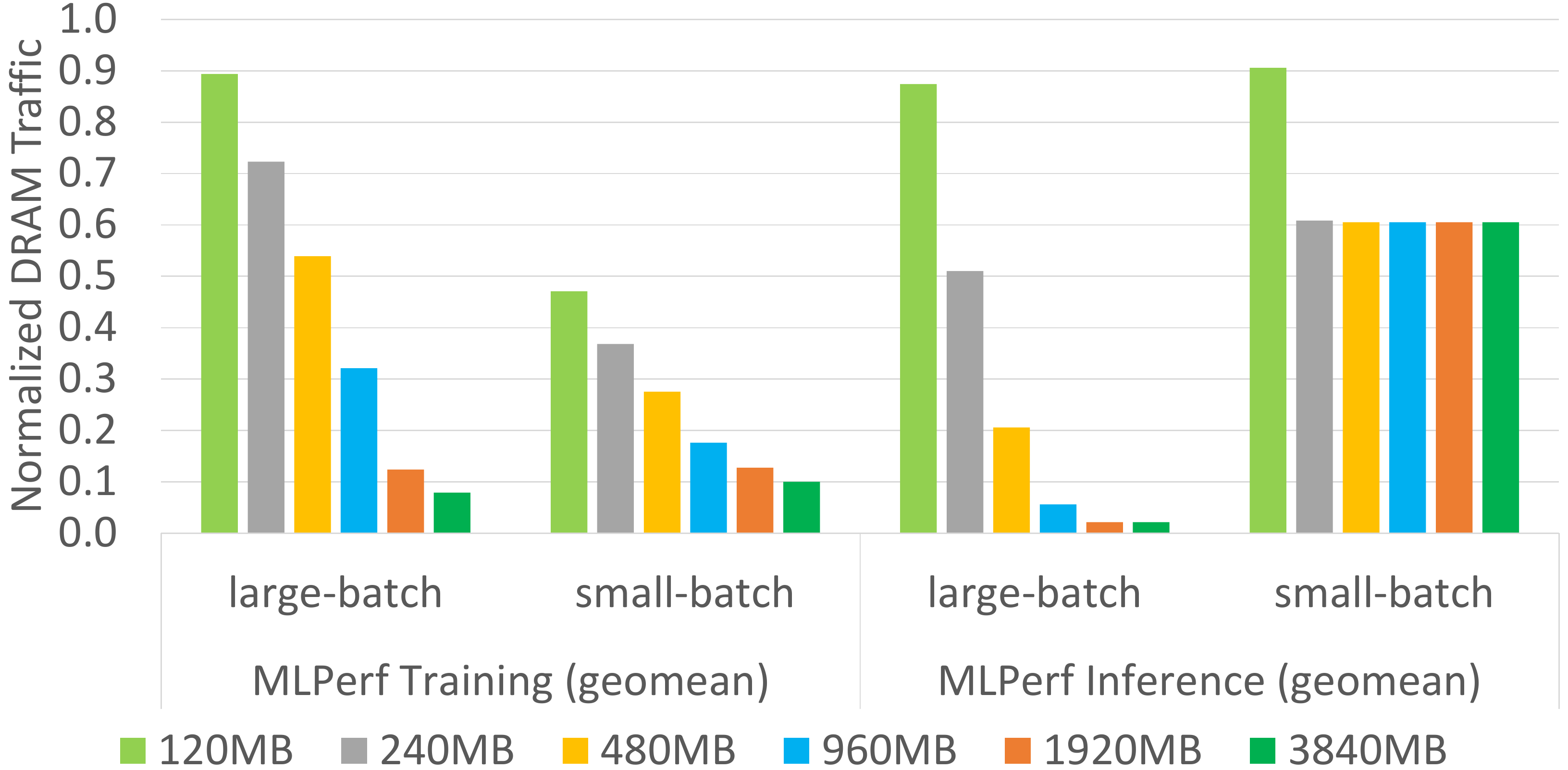}
\caption{DRAM traffic reduction versus LLC capacity, normalized to the baseline GPU-N configuration with 60MB of LLC.}
\label{fig:dram_traffic}
\vspace{-0.2in}
\end{figure}

\vspace{-0.05in}
\subsection{Very Large Caches for DL-optimized GPUs}
\label{sec:sec:very_large_caches_for_gpus}

Because the memory bandwidth demands of DL applications are likely to
exceed what evolutionary DRAM scaling can provide, GPU designers must
pursue alternative methods to meet aggressive bandwidth targets.
Historically, the GPU's LLC has remained relatively small because HPC
workloads have good spatial locality and the majority of the
off-chip bandwidth filtering potential can be captured within last level caches
that are measured in tens of megabytes. In DL workloads data locality
spans multiple temporal and spatial scales, requiring much larger
capacities. 

To understand how much LLC capacity GPU-N would need to
effectively shield the GPU's DRAM system, we examine the off-chip DRAM
traffic reduction achieved when sweeping GPU-N's LLC capacity from
60MB to 4GB\@.
Figure~\ref{fig:dram_traffic} shows that doubling the LLC capacity to
120MB provides up to 53\% reduction in off-chip DRAM traffic in DL
training.  Further growth to 960MB reduces off-chip BW demand by 82\%
(5$\times$ reduction). We will later show that DRAM traffic reduction from larger caches
correlates well with  improved training performance. 
Compared to DL training,
large LLC capacities are even more beneficial for DL inference because they
enable a larger portion of the workload's weights and activations to
be cached on-chip. For example, in large-batch inference, a 960MB LLC achieves a
16$\times$ DRAM traffic reduction, while a 240MB LLC is sufficient to
capture all available data reuse data on-chip for small-batch inference because of
its smaller memory footprint. 

Because modern GPUs are already area limited~\cite{MCM-GPU},
implementing hundreds of additional megabytes of on-die LLC without
severely sacrificing other functionalities is not feasible. Even if it
could be done, such an architecture would be out of balance
for optimal HPC performance.  Because of the different sensitivities
to off-chip GPU bandwidth and on-GPU cache capacities between HPC and
DL workloads, we now explore the potential benefits of a composable GPU
architecture that can satisfy the unique demands of each domain while
maximizing design-reuse through a modular on-package design.

\vspace{-0.05in}
\section{Composable GPU Architecture}
\label{sec:arch}
A COPA-GPU architecture has two primary goals: (i) to largely preserve the
existing GPU architecture to minimize design effort and maximize
reuse, and (ii) to provide flexibility in specializing the GPU memory
system for diverging requirements across application
domains. Specifically, we aim to improve the GPU's memory system
with substantially more cache and memory bandwidth
than is required for domains such as HPC\@, to unlock deep learning performance on GPUs.  To
achieve this range of GPU system capabilities, we leverage
on-package multi-chip-module integration to couple a GPU-core die
with different memory system dies, each having different allocations of
on-package cache and memory resources.


\begin{figure}[t]
  \center
  \includegraphics[width=1\linewidth,trim= 12 0 50 0,clip]{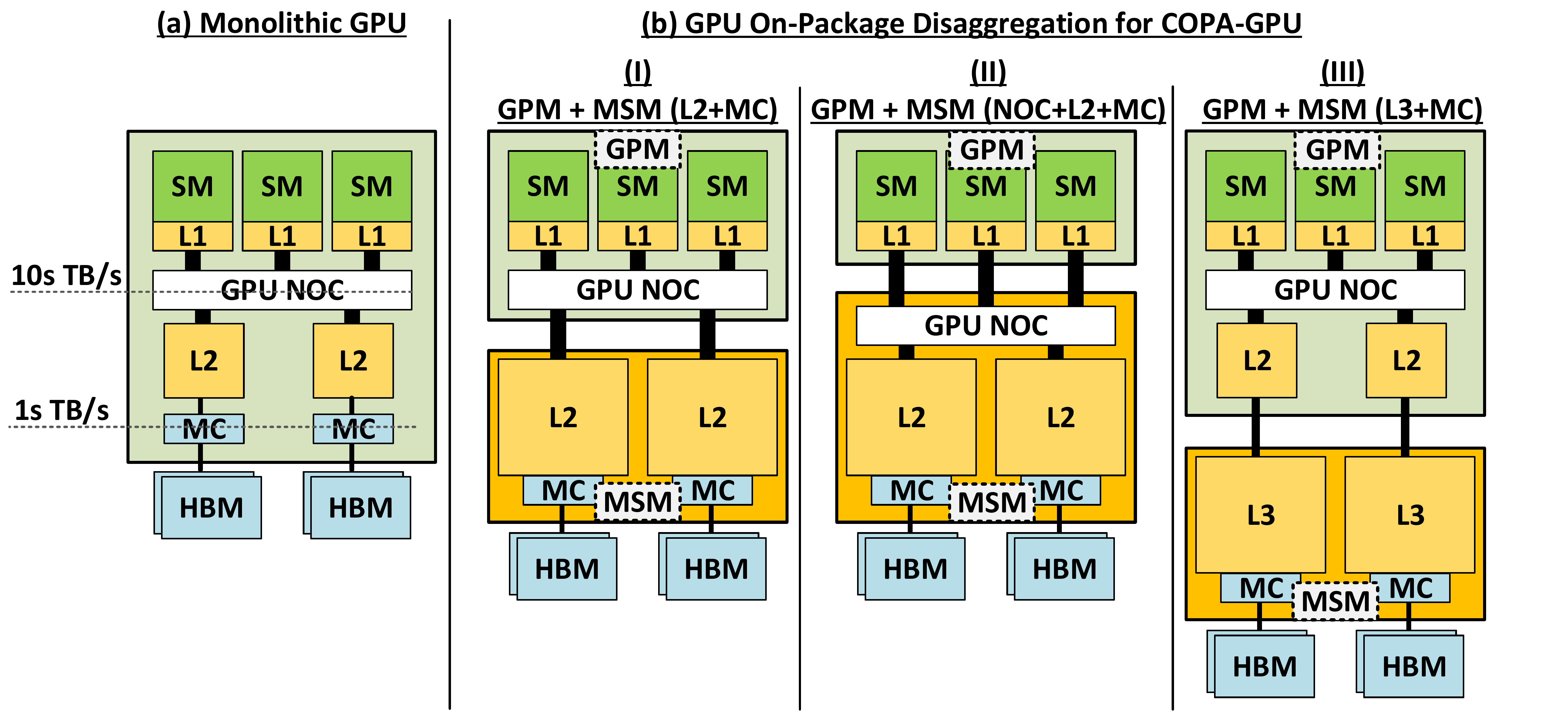}
  \caption{GPU MCM disaggregation options for practical and composable domain specialization.}
  \label{fig:copa-disaggregation}
  \vspace{-0.2in}
\end{figure}

\begin{figure*}[ht]
  \center
  \includegraphics[width=1\linewidth, trim= 0 10 5 5,clip]{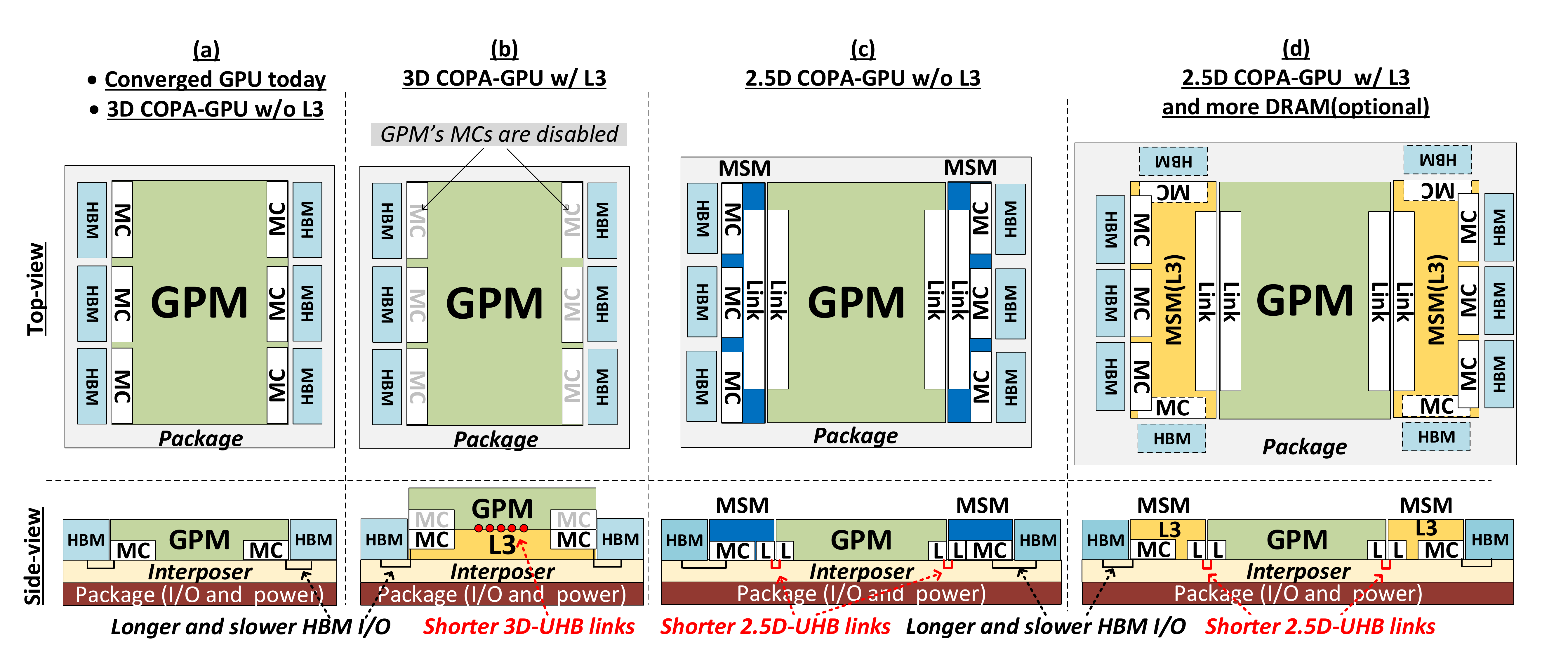}
  \caption{COPA-GPU architectural options spanning 2.5D and 3D integration domains.}
  \label{fig:copa-gpu-diagrams}
  \vspace{-0.2in}
\end{figure*}


\subsection{Practical GPU Disaggregation With COPA-GPU}
We propose to disaggregate today's monolithic GPU into a basic GPU
Module (GPM) on-package that is reused without modifications across
domain-specialized COPA-GPU instances and a domain-specialized Memory
System Module (MSMs) tailored for the specifics of each application
domain. The main challenge with on-package disaggregation for GPUs is
striking the right balance between the degree of composability and the
practicality of effectively shifting the burden of high bandwidth
intra-GPU communication from the on-chip to on-package wires.
 
Figure~\ref{fig:copa-disaggregation}(a) depicts a generic monolithic
GPU architecture, with streaming multiprocessors (SMs) and
corresponding L1 caches connected to a distributed L2 cache via the
GPU network on chip (NoC). L2 cache slices are attached to Memory Controllers (MCs)
driving the off-chip DRAM memory interface. While contemporary
off-chip DRAM interfaces provide a few TB/s of memory bandwidth (2.7
GB/s in GPU-N), the on-chip interconnects of a modern GPU are designed
to transport many tens of TB/s of on-chip bandwidth between the SMs
and the on-chip memory hierarchy.

The three possible GPU disaggregation options are shown in
Figure~\ref{fig:copa-disaggregation}(b)\@. Option (I) denoted by
GPM+MSM(L2+MC) features a GPM containing SMs, L1s, and the GPU NoC,
while the MSM includes the L2 cache and the memory controllers. This
design can provide more effective die area and die edge than single
reticle-limited die to support bigger L2 cache and higher DRAM
bandwidth when needed.  Option (II) denoted by GPM+MSM(NoC+L2+MC) is
similar to (I), but shifts the GPU NoC from the GPM onto the MSM\@.
However, options (I) and (II) are impractical as they both require many
tens of TB/s of NoC traffic to instead traverse on-package wires,
which Section~\ref{sec:sec:copa-gpu-technologies} later shows is not
realizable using current or known proposed technologies.

Thus, we choose option (III) denoted by GPM+MSM(L3+MC) for exploring a
practical COPA-GPU design as it relies on existing L2 cache bandwidth
filtering within the GPM to achieve feasible levels of off-GPM
inter-die bandwidth. Additionally, propose to add an additional layer of L3 cache
between the L2 cache and MC, and to have both the L3 cache and the MC
implemented on the MSM\@. Such composable design re-organization can 
provide more than 4$\times$ higher post-L2 bandwidth using previously proposed
package integration technologies.


\subsection{COPA-GPU Architectures and Packaging}
\label{sec:sec:practical_copa}


 


The architectural domain customization in our proposed
COPA-GPU is achieved through integration of a GPM with a dedicated
domain-optimized MSM using 2.5D or 3D on-package integration, that could
leverage either planar or vertical die stacking approaches. We
consider COPA-GPU architectural options spanning both
integration strategies while taking into account the unique
interdependency between architectural and package technology choices,
distilling the advantages and disadvantages of each approach.

Figure~\ref{fig:copa-gpu-diagrams}(a) shows a high-level top and side
view of a modern GPU architecture, including a maximum-sized GPU die
with the L2 cache and memory controllers on-die and HBM sites attached
via 2.5D on-silicon interposer integration. The same diagram shown in
Figure~\ref{fig:copa-gpu-diagrams}(a) also describes the basic
COPA-GPU GPM, pre-equipped to be 3D integrated with an MSM die to
provide a 3D DL-optimized COPA-GPU variant. 

Figure~\ref{fig:copa-gpu-diagrams}(b) illustrates a 3D
organization of a DL-optimized COPA-GPU where the GPM die from
Figure~\ref{fig:copa-gpu-diagrams}(a) is integrated with an MSM
carrying additional L3 cache. The MSM is vertically attached via a 3D
ultra-high bandwidth (3D-UHB) link using high-density intra-die
bonding, with up to 14.7 TB/s of bandwidth, further detailed in
Section~\ref{sec:sec:copa-gpu-technologies}\@. The MSM is
positioned between the GPU and silicon interposer, and provides all
the essential connectivity between the GPM and the silicon interposer
via through silicon vias (TSVs), as shown in
Figure~\ref{fig:copa-gpu-diagrams}(b:side-view). The composable nature
of a 3D COPA-GPU allows for designs both with (for DL) and without
(for HPC) additional L3. The primary disadvantage of a 3D COPA-GPU
organization is that it cannot provide any additional die edge to
improve DRAM scaling.

Figure~\ref{fig:copa-gpu-diagrams}(c) shows a basic 2.5D COPA-GPU
integrating a GPM with up to two MSM modules in a 2.5D organization
targeting HPC\@. The design of a 2.5D COPA-GPU retains most of its
original functionality on the base GPM but offloads the memory
controllers (MC) and HBM I/O to new (small) in-package MSMs. A 2.5D
COPA-GPU has the advantage that offloading of GPM area allocated to MC
and HBM I/O frees up die area for the implementation of UHB links and
additional compute resources in the baseline COPA-GPU.

Figure~\ref{fig:copa-gpu-diagrams}(d) augments the 2.5D COPA-GPU in
Figure~\ref{fig:copa-gpu-diagrams}(c) with a large L3 cache and
additional DRAM stacks, subject to package area limitations. The same
GPM die is reused between Figure~\ref{fig:copa-gpu-diagrams}(c) and
(d) and this 2.5D DL-optimized organization has several advantages compared to a 3D organization.
First, it can provide up to 2$\times$ more L3 capacity than the 3D
organization. Second, the additional MSM geometry increases the total
available die-edge of the design. This die-edge can provide up to 1.7$\times$ or 2.3$\times$ higher HBM BW and larger
capacity, via 10 total HBM sites, or 14 total HBM sites using
maximally sized MSMs (not shown).


\vspace{-0.05in}

\subsection{COPA-GPU Microarchitectures}
\label{sec:sec:uarch}

\begin{figure}[t]
  \center
  \includegraphics[width=1\linewidth]{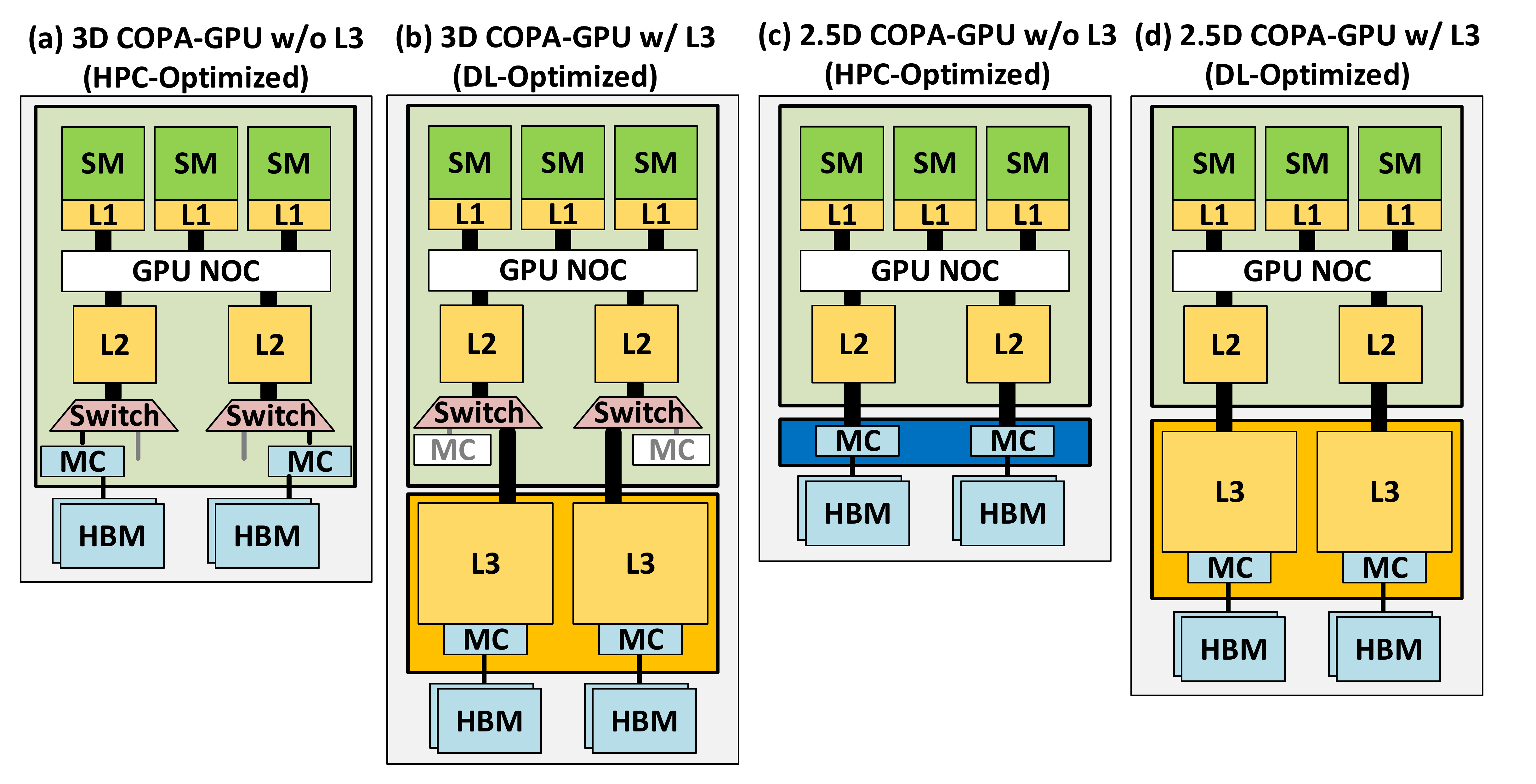}
  \caption{L3 cache microarchitecture in COPA-GPU designs.}
  \label{fig:copa_uarch}
  \vspace{-0.2in}
\end{figure}

Figure~\ref{fig:copa_uarch} presents the microarchitecture of the
proposed memory systems for both the 3D and 2.5D COPA-GPU designs.
Figures~\ref{fig:copa_uarch}(a) and (b) outline the main components of
a 3D COPA-GPU with and without an extended L3 MSM\@. The most
significant difference between current GPU architectures and the
3D COPA-GPU is the addition of a new switching component that
steers requests from the L2 to the on-chip memory controller
or to the UHB links, depending on whether the MSM is present in-package.

If the MSM is not present (Figure~\ref{fig:copa_uarch}(a)), the switch
is configured to steer the memory requests directly from the L2 to its local
MC, similarly to today's GPUs\@. If the MSM is present,
(Figure~\ref{fig:copa_uarch}(b)), the switch is reconfigured to steer
post-L2 traffic to the L3 via the UHB link. This design requires the
MSM to implement its own MC that connects to HBM I/O, with silicon
area overheads stemming from adding the UHB link, the TSV-based I/O,
and power delivery between GPM and the silicon interposer dies. In the
2.5D COPA-GPU designs without and with additional L3 capacity on the
MSM (Figures~\ref{fig:copa_uarch}(c) and (d)), the GPM is completely
stripped of its original MC and HBM I/O, and post-L2 traffic is always
routed to the MSM modules over on-package UHB links. Both the 2.5D
COPA-GPU configurations consist of the same GPM, but are equipped with
different versions of the MSM to serve different application classes 
(i.e. with or without L3 and additional HBM sites).

In both designs, the L3 cache is architecturally implemented as yet another level of
memory-side cache that (when present) backs the existing L2\@. It is
neither inclusive nor exclusive, nor does it require coherence with the L2
because the L2 already serves as the point of coherence in GPU
systems.  Lines present in the L2 always supercede lines in the L3
from the GPU's perspective and are written back to the L3 upon
eviction from the L2\@. No requests are routed to the L3 without first
being serviced at the L2\@.

\vspace{-0.05in}
\subsection{Costs and Benefits of the COPA-GPU}
\label{sec:sec:cost}

The major drawback of a 2.5D COPA-GPU organization is that it
increases the package size. Whereas the 3D COPA organization has minimal
impact on package complexity (same top view), but the base GPM must
account for the implementation of the distributed on-die UHB link
bonding used for vertical inter-die communication
(Section~\ref{sec:sec:copa-gpu-technologies}) and the TSVs overheads
used for traversing through the MSM die. We estimate that in the 3D
case, the cost of providing as much as 14.7 TB/s bandwidth of UHB I/O
will consume less than 4\% of silicon surface and metal layers for
inter-die communication bonding. The majority of the active area and
lower-level metals under the 3D bonding can likely be reused for other
logic in the GPM\@. In 2.5D designs, the maximally sized links results in
approximately 6\% area overheads assuming a 20Gbps signaling rate.
Section~\ref{sec:sec:copa-gpu-technologies} provides further details
on these assumptions.

Even though the COPA-GPU approach introduces 4\%-6\% area overhead due
to new MCM communication interfaces, this area (and thus cost overhead) will be more
than offset by the savings in the HPC oriented variants by not
carrying forward the unnecessary and expensive DL-oriented memory
subsystem. Moreover, we believe that in the future even non-composable
GPU designs will turn to MCM organizations due to looming reticle limitations,
effectively introducing similar MCM link overheads into \textit{all} GPU designs.

Adding new communication interfaces that memory system requests must
traverse comes with an additional energy tax. We estimate that the
2.5D organized UHB-link consumes less than 9 Watts, assuming peak
100\% bandwidth utilization at 0.3pJ/b with a 25\% wire toggle-rate.
The 3D link consumes less than 2 watts, due to its more efficient link
technology and overall shorter communication distances
(Section~\ref{sec:sec:copa-gpu-technologies}).

A large L3 reduces the number of DRAM accesses, more than offsetting
the additional cost of the UHB link traversal and L3 accesses.
Figure~\ref{fig:dram_traffic} shows that the 960MB and 1.9GB L3
configurations reduce overall DRAM traffic by up to 94\% and 98\%,
respectively. We estimate that fetching a cache line from an
SRAM-based COPA-GPU L3 into the GPM will consume approximately
4$\times$ less energy than accessing HBM memories. These estimates
fully account for the energy associated with traveling to an SRAM
sub-array on a MSM-die and back, as well as the energy consumed by the
SRAM sub-array~\cite{EnergyMCM,keckler11,EfficientMemory,SRAM17}\@.
Consequently, we estimate that either COPA-GPU design (utilizing a
960MB L3 cache) will reduce the total HBM-related GPU energy
consumption by up to 3.4$\times$\@. While left for future exploration,
the cost of the MSM caching modules could be reduced by implementing
them in older and less expensive processes.


\vspace{-0.05in}
\subsection{COPA-GPU Enabling Technologies}
\label{sec:sec:copa-gpu-technologies}
The feasibility of COPA-GPUs depends on several technologies reaching maturity
for industrial implementation.

\begin{table}[t]
  \centering
      \caption{Bandwidth and energy characteristics assumptions for 2.5D and 3D ultra-high bandwidth (UHB) in this work.}
    \label{tab:uhb_link}
  \small
      \begin{tabular}{|c|c|c|c|c|c|c|}
      \hline
      Technology & BW Density      & Max Bisection BW & Energy/bit  \\ \hline \hline
      2.5D       & 256GB/s/mm     & 14.7TB          & 0.3 pJ/b   \\ \hline
      3D         & 512GB/s/mm$^2$ & $>$14.7TB         & 0.05 pJ/b   \\ \hline
      \end{tabular}
\end{table}

{\bf Cache Technology Projections:} Large caches can be realized
through high density Embedded DRAM (eDRAM)~\cite{EDRAM-IBM-ISSCC20} or
SRAM technologies~\cite{SRAM-INTEL-ISSCC18,SRAM-TSMC-ISSCC20}\@. IBM
recently implemented a 960MB cache on a 696mm$^2$ die using a 14nm
eDRAM technology~\cite{EDRAM-IBM-ISSCC20}\@. Graphcore's recently
announced second-generation IPU~\cite{graphcore_2020} integrates 900MB
of SRAM along with thousands of cores on a single 823 mm$^2$ die using
TSMC's 7nm process. Cache density is dictated by combination of the
bitcell area, control overheads, and the bandwidth
requirements. Because the COPA-GPU's L3 is designed to provide lower
bandwidth than L2, we estimate that a future reticle limited 826
mm$^2$ MSM die (the same die size as an NVIDIA A100 GPU~\cite{ampere}),
can provide up to 2GB of L3 cache. Though, the remainder of this work assumes
a conservative projection of a 960MB L3 on an 826mm$^2$ die, which implies
a maximum of 960MB L3 in a 3D COPA-GPU with single MSM die and 1920MB L3 in a 2.5D
COPA-GPU with two MSM dies.

{\bf High Bandwidth 2.5D and 3D Interconnects:} High-speed links that enable
2.5D integration are rapidly maturing~\cite{Turner-GRS, INTELEMIB,
USR,ARMCOWOS}\@. Chen et al.~\cite{USR} recently demonstrated 20Gbps signaling rates
across a 2.5mm silicon interposer layer at 0.3pJ/b, resulting in a
$\sim$200GB/s/mm bandwidth density per layer, which can be further increased at
shorter distances. This work assumes a 2.5D COPA-GPU UHB link with a bandwidth
density of 256GB/s/mm\@, thus the dedicated edges of the 826mm$^2$ GPM module can
provide up to 14.7TB/s of off-GPM bandwidth as described in
Figure~\ref{fig:copa-gpu-diagrams}(d)\@. While this bandwidth to the MSM far
exceeds the actual L3 bandwidth requirement, it provides headroom for future
increases in off-die communication.

3D integration is also
maturing~\cite{Lakefield,Foveros,CAE-LETI-3D,SOIC, SOIC1}\@. For
example, TSMC's System on Integrated Chips (SoIC) is expected to
provide ultra-dense 3D interconnects~\cite{SOIC, SOIC1} with greater
than 1 TB/s/mm$^2$ of inter-die bandwidth, assuming a 1 Gbps signaling
rate. We conservatively assume 3D-UHB links with 512GB/s/mm$^2$ of
bandwidth density at 0.05 pJ/b\@. Achieving the assumed 14.7TB/s of data
bandwidth for a 2.5D COPA-GPU would require 28.7mm$^2$ (less than 4\%)
of silicon area for the inter-die communication
bonding. Table~\ref{tab:uhb_link} summarizes the UHB link
characteristics for both 2.5D and 3D scenarios.

\vspace{-0.05in}
\section{Evaluation}
\label{sec:eval}
\begin{table}[t]
  \centering
  \caption{MLPerf training and inference benchmarks.}
  \label{tab:mlperf}
  \small
  \renewcommand\tabcolsep{3pt}
  \begin{tabular}{|l|r|r|r|r|r|}
  \hline
                 &         & \multicolumn {2}{c|} {Small-batch} & \multicolumn {2}{c|} {Large-batch} \\
  \cline{3-6} 
  Benchmark    & \multicolumn{1}{c|}{Type} & Per-GPU    &  \multicolumn{1}{c|}{Memory}    & Per-GPU       &  \multicolumn{1}{c|}{Memory}             \\ 
  &              & batch size &  \multicolumn{1}{c|}{footprint} & batch size    &  \multicolumn{1}{c|}{footprint}          \\ \hline
  resnet         & training  & 12         & 989MB            & 128    & 6GB       \\\hline
  ssd            & training   & 4          & 559MB            & 128    & 7.9GB     \\\hline
  maskrcnn       & training  & 1          & 2.1GB            & 6      & 9.9GB     \\\hline
  minigo         & training    & 128        & 105MB            & 2,048  & 1.5GB     \\\hline
  gnmt           & training  & 32         & 3GB              & 256    & 8.3GB     \\\hline
  transformer    & training  & 640        & 4.5GB            & 5120   & 7.9GB     \\\hline
  ncf            & training  & 65,526     & 657MB            & 1,048,576 & 4.5GB  \\\hline
  resnet         & inference  & 1     &  49MB            & 232 & 1.1GB      \\\hline
  mobilenet      & inference  & 1     &  16MB            & 704 & 2GB     \\\hline
  ssd-small      & inference  & 1     &  24MB            & 288 & 2GB     \\\hline
  ssd-large      & inference  & 1     &  136MB            & 6 &  562MB     \\\hline
  gnmt           & inference  & 1     &  300MB             & 128 & 961MB     \\\hline
  \end{tabular}
\vspace{-0.1in}
\end{table}

We now present our simulation methodology, examine the performance sensitivity
of the baseline COPA-GPU architecture to DRAM bandwidth and on-package LLC capacity,
and evaluate several specific DL-specialized COPA-GPU configurations in both
training and inference scenarios.

\subsection{Methodology}
\label{sec:sec:methodology}

To provide performance projections for DL workloads, we perform our
studies using unmodified workloads from the MLPerf training and
inference benchmark suites~\cite{mlperf_mlsys_2020, mlperf_isca_2020}
for one end-to-end iteration of the workloads, rather than using
disjoint and isolated GPU kernel calls. This approach allows us to
characterize the overall throughput and capture a new important class
of inter-kernel data reuse that will drive future memory system design
on GPUs\@. MLPerf is the de facto standard for DL benchmarking
and is maintained by major DL chip vendors including NVIDIA,
Google, and Intel\@. Table~\ref{tab:mlperf} shows the seven MLPerf
training benchmarks and the five inference benchmarks used, taken from
NVIDIA's publicly available MLPerf training v0.6 and MLPerf inference
v0.5 submissions~\cite{mlperf_training_2019, mlperf_inference_2019}.
These codes are highly optimized for NVIDIA GPUs by exploiting
high-throughput Tensor Cores\mbox{\cite{volta}} and have demonstrated
performance scalability from a single to thousands of GPUs.

\begin{table}[t]
  \centering
  \caption{Detailed GPU configurations.}
  \label{tab:config}
  \small
  \renewcommand\tabcolsep{3pt}
  \begin{tabular}{|l|c|c|c|}
  \hline
  Configuration & NVIDIA V100 & NVIDIA A100 & GPU-N        \\ \hline
  SMs                  & 80     & 108        & 134          \\\hline
  GPU frequency (GHz)  & 1.4    & 1.4        & 1.4         \\\hline
  FP32 (TFLOPS)        & 15.7   & 19.5       & 24.2         \\\hline
  FP16 (TFLOPS)        & 125    & 312        & 779        \\\hline
  L2 cache (MB)        & 6      & 40         & 60         \\\hline
  DRAM BW (GB/s)       & 900    & 1,555      & 2,687        \\\hline
  DRAM Capacity (GB)   & 16     & 40         & 100        \\\hline
  \end{tabular}
\vspace{-0.1in}
\end{table}

\begin{figure*}[t]
  \begin{subfigure}[t]{1\textwidth}
    \center
    \includegraphics[width=1\linewidth]{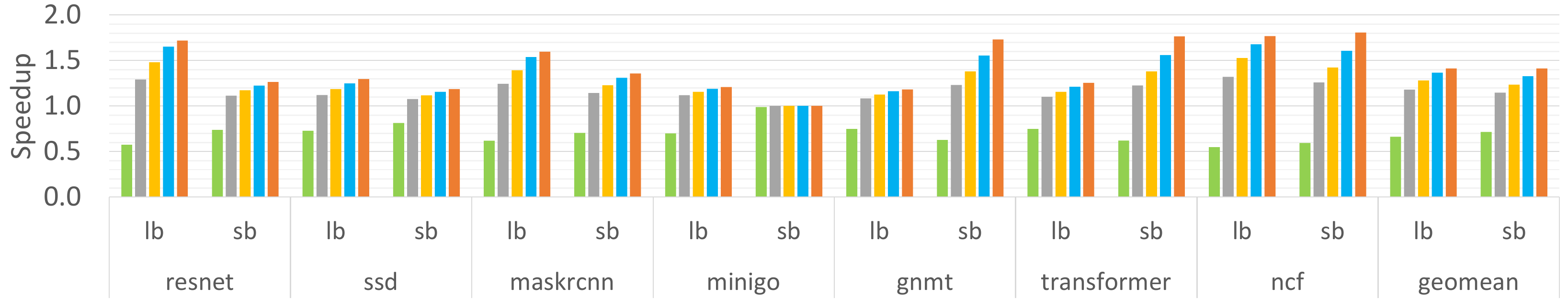}
    \caption{Training}
    \label{fig:dram_bw_train}
  \end{subfigure}
  \begin{subfigure}[t]{1\textwidth}
    \center
    \includegraphics[width=1\linewidth]{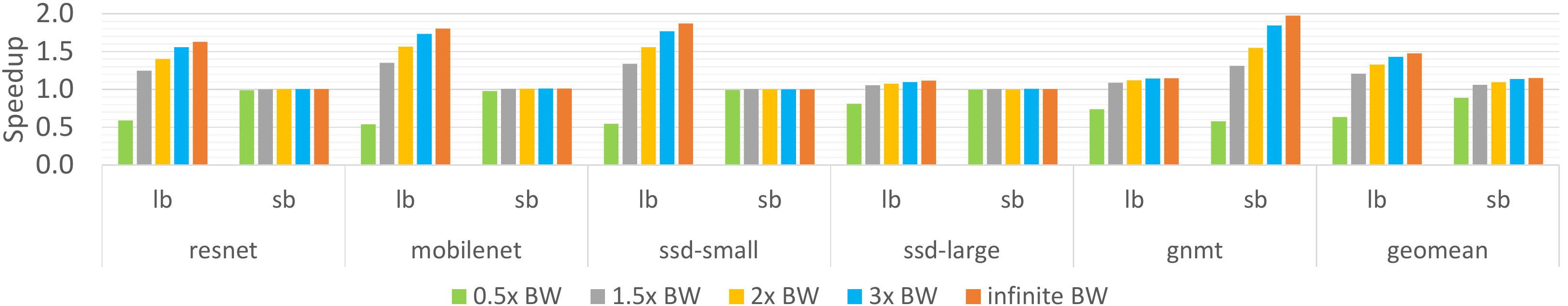}
    \caption{Inference}
    \label{fig:dram_bw_inference}
  \end{subfigure}
    \caption{Performance of a basic COPA-GPU with varying DRAM bandwidth for large-batch (\textit{lb}) and small-batch (\textit{sb}) settings, normalized to the baseline GPU-N performance.}
  \label{fig:dram_bw}
  \vspace{-0.2in}
\end{figure*}

To build a complete picture of future deep learning scenarios, we
run all our DL benchmarks in two different configurations. For DL
training, we use a large per-GPU batch size to characterize a
single-GPU training situation and a small per-GPU batch size to represent a
large-scale training system. For DL inference, we use a large per-GPU
batch size to represent a datacenter processing inference task with
large numbers of concurrent queries and a small per-GPU batch size for
an edge-device scenario with real-time processing
requirements. Although small-batch inference workloads do not commonly
run on the large datacenter-grade GPUs we evaluate in this paper, we
include them in our studies for completeness as future GPU designs
could also be scaled down (in aggregate) to build niche-specific GPU
products in traditional or COPA-GPU form.

The batch sizes we choose for each scenario are taken from NVIDIA's
MLPerf submissions and are shown in
Table~\ref{tab:mlperf}\@. For our evaluations, we execute each
application on a single NVIDIA Tesla V100 and collect a GPU execution
trace from a full end-to-end iteration. We focus on the per-GPU
workload analysis and omit the all-reduce synchronization overheads
when projecting performance for large-scale DL training.  All-reduce
performance largely depends on the implementation of the inter-GPU
network, which is beyond the scope of this work but is also receiving
attention within the architecture community~\cite{inswitch_isca_2019,
innetwork_isca_2020}.

We simulate these workloads using a trace based, partially execution
driven, architectural-level GPU simulator using methodology similar to
that of NVIDIA's NVArchSim~\cite{nvas}\@ and AccelSim~\cite{AccelSim}.
Our simulator provides very fast simulation of deterministic portions
of an application without sacrificing simulation accuracy or
correctness during non-deterministic application phases.  Our
simulator has been correlated to NVIDIA Tesla V100 GPU using hundreds
of HPC and ML/DL workloads and achieves a 0.986 Pearson correlation to
real hardware.

Our analysis focuses on the projected GPU-N configuration described in
Table~\ref{tab:config}\@. GPU-N's SM count growth beyond NVIDIA
A100's is proportional to the projected FP32 throughput (from
Table~\ref{tab:gpu_trend}) with a constant SM frequency.  
GPU-N's DRAM capacity is projected based on a linear
extrapolation similar to the projections done for the DRAM bandwidth.

\begin{figure*}[t]
  \begin{subfigure}[t]{1\textwidth}
    \center
    \includegraphics[width=1\linewidth]{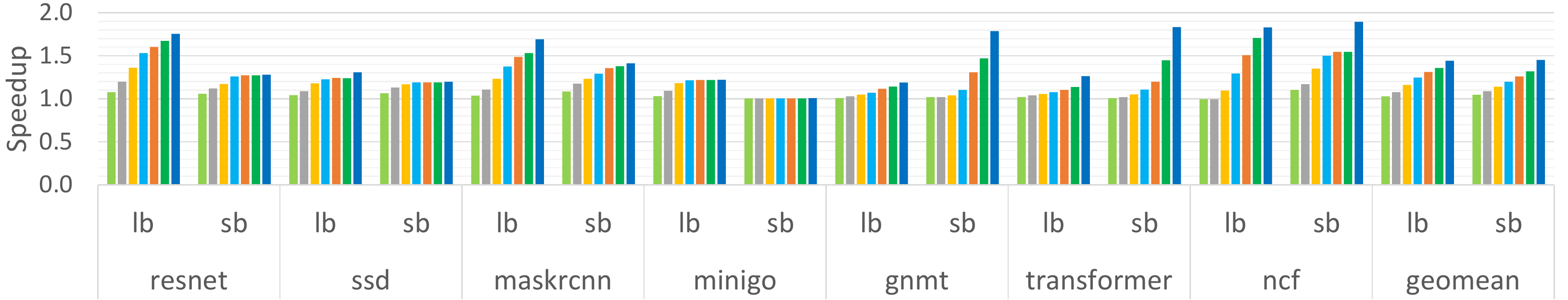}
    \caption{Training}
    \label{fig:l2size_train}
  \end{subfigure}
  \begin{subfigure}[t]{1\textwidth}
    \center
    \includegraphics[width=1\linewidth]{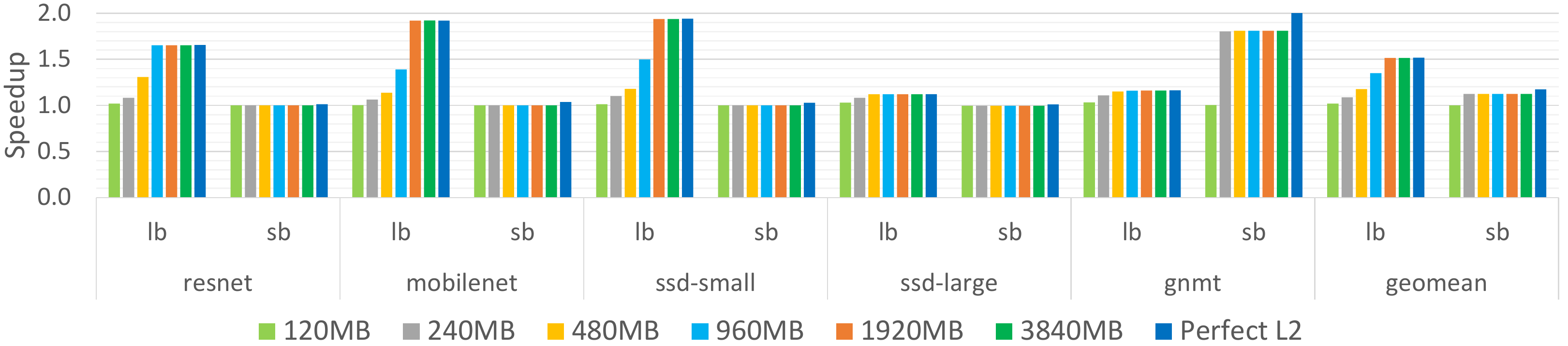}
    \caption{Inference}
    \label{fig:l2size_inference}
  \end{subfigure}
    \caption{Performance of a basic COPA-GPU with varying LLC capacities for both large-batch (\textit{lb}) and small-batch (\textit{sb}) settings, normalized to the baseline GPU-N performance.}
  \label{fig:l2size}
  \vspace{-0.2in}
\end{figure*}

\vspace {-0.05in}
\subsection{DL Performance Sensitivity to DRAM Bandwidth}
\label{sec:sec:dram_bw_sweep}

We perform a detailed sweep of the DRAM bandwidth settings for the
basic COPA-GPU design without L3 (similar to GPU-N) from half of its nominal
bandwidth (1.3TB/s) up through infinite bandwidth.
Figure~\ref{fig:dram_bw} summarizes the overall performance normalized
to the 2.7TB/s baseline. Workload performance scales steadily with 
growing DRAM bandwidth up to a 3$\times$BW
(8.1TB/s) setting in most of the training and inference scenarios,
with diminishing returns beyond this point. Small-batch inference
applications are less sensitive to DRAM bandwidth due to their
relatively small footprint (more details in
Section~\ref{sec:sec:LLC_sweep}), while the remaining DL training and
inference workloads show that even a relatively modest 1.5$\times$
increase in DRAM bandwidth from 2.7TB/s to 4TB/s would lead to notable
speedups of up to 18\% for training and 21\% for inference.

From these results we conclude that modern GPUs are on track to be severely DRAM
bandwidth limited and that additional raw DRAM bandwidth is
the most powerful tool for scaling DL performance. However, because
evolutionary improvement in HBM memory frequency and pin density 
is unlikely to provide a substantially higher
bandwidth scaling ratio than our aggressively projected 1.7$\times$ (for
GPU-N), we explore the potential of a very large
and high bandwidth on-package cache to close this gap in future COPA-GPUs.

\vspace{-0.05in}
\subsection{DL Performance Sensitivity to Cache Capacity}
\label{sec:sec:LLC_sweep}
To understand the impact of large caches on DRAM bandwidth reduction in future
GPUs, we sweep the LLC (L2) cache capacity of the basic COPA-GPU without L3
from 60MB to 3,840MB to characterize
the performance potential of the larger on-package caches. 
A perfect L2 where all requests hit in the L2
caches serves as the upper bound on performance. Figure~\ref{fig:l2size}
shows the performance of these different L2 configurations, normalized to
the 60MB baseline. Increasing the L2 capacity results in performance gain
equivalent to large increases in DRAM bandwidth, primarily because of the significant
off-chip DRAM traffic reduction, as shown in Figure~\ref{fig:dram_traffic}.

For DL training, 480MB of L2 performs slightly worse than a 1.5$\times$
increase in DRAM bandwidth and 960MB of L2 performs slightly worse than a 2$\times$
increase in DRAM bandwidth (comparing to Figure~\ref{fig:dram_bw}). Yet larger L2
caches continue to incrementally improve the performance, but even a 3,840MB L2 still
results in an 8\% and 13\% performance gap in large-batch and
small-batch DL training workloads when compared to a perfect L2.
Because DL inference applications generally have smaller memory
footprint sizes than DL training, performance saturates once all workload data can be
cached. The saturation points are 1,920MB and 240MB for large- and  
for small-batch inference respectively, which correspond to their
memory footprints listed in Table~\ref{tab:mlperf}.

We conclude that a substantially larger LLC is an
attractive solution for scaling DL performance within a COPA-GPU\@, however
closing the performance gap to a perfect L2 would require 4GB
of on-package cache, which is impractical even under aggressive technology projections. 
As a result, to maximize DL performance COPA-GPUs must not just utilize a very large L3 cache, 
but combine a large L3 and higher DRAM bandwidth made
available with additional MSM-die edge in a 2.5D COPA-GPU design.

\begin{table}[t]
  \centering
 \caption{COPA-GPU architectural parameters.}
  \label{tab:arch}
  \small
  \renewcommand\tabcolsep{3pt}
      \begin{tabular}{|l|c|c|c|}
       \hline
      Architecture   & LLC capacity & DRAM BW & DRAM capacity \\ 
      configuration  & (MB) & (TB/s) & (GB) \\\hline
      GPU-N       & 60                & 2.7              & 100    \\ \hline
      HBM+L3        & 960               & 2.7              & 100    \\ \hline
      HBML+L3       & 960               & 4.5              & 167    \\ \hline
      HBM+L3L       & 1920              & 2.7              & 100    \\ \hline
      HBML+L3L      & 1920              & 4.5              & 167    \\ \hline
      HBMLL+L3L     & 1920              & 6.3              & 233    \\ \hline
      Perfect L2   & infinite          & infinite         & infinite  \\ \hline
      \end{tabular}
\end{table}

\subsection{DL Performance Scaling with COPA-GPU}
\label{sec:sec:dl_perf}

Table~\ref{tab:arch} summarizes the DL-optimized COPA-GPU
configurations that are enabled by architectural choices outlined in
Section~\ref{sec:arch}\@. The HBM+L3 design provides 960MB of L3 cache
through either 3D stacking with a single maximum sized MSM die, or
2.5D stacking with two MSM dies, each set to half of the maximum die
size. The option with more HBM resources, denoted by an HBML in the
HBML+L3, applies only to 2.5D, by exploiting the extra MSM die edge
area to provide a 1.7$\times$ higher HBM bandwidth and a larger HBM
capacity (10 HBM sites in total). With two maximally-sized MSM dies in
2.5D stacking scenario, the L3 cache capacity and the additional DRAM
bandwidth can be scaled to up to 1,920MB and 2.3$\times$ of the
baseline DRAM bandwidth (via 14 HBM sites) respectively. As a result, we
consider three additional COPA-GPU configurations with 1,920MB of L3
cache and varying HBM resources: HBM+L3L, HBML+L3L and HBMLL+L3L, with
2.7TB/s, 4.5TB/s and 6.3TB/s of DRAM bandwidth, respectively.

\begin{figure}[t]
  \center
  \includegraphics[width=1\linewidth]{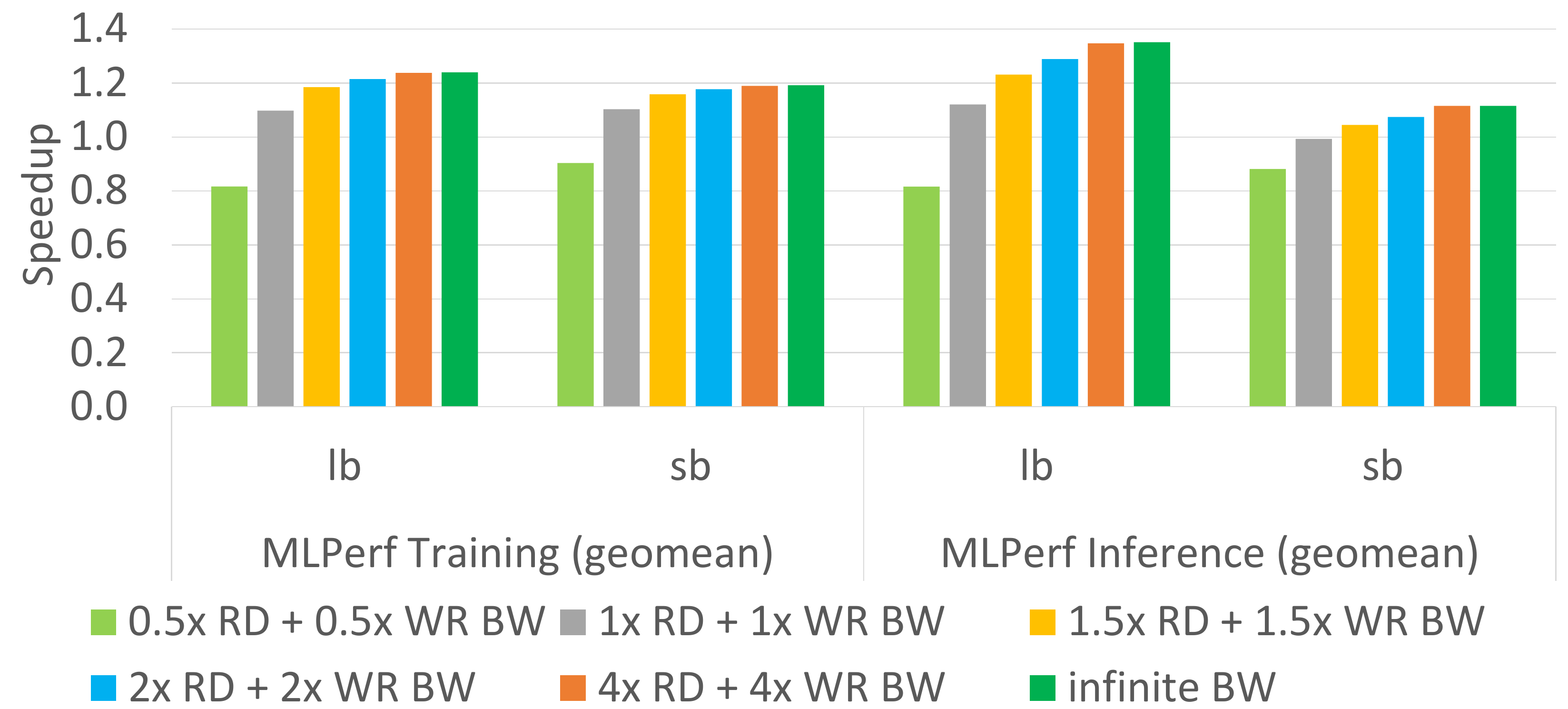}
\caption{Performance of a HBM+L3 COPA-GPU with varying UHB link bandwidth to the
L3 cache, normalized to the baseline GPU-N performance.}
\label{fig:l3_bw}
\vspace{-0.2in}
\end{figure}

Before moving to overall results, we first analyze the UHB on-package link bandwidth requirements by
sweeping the unidirectional L3 bandwidth in the HBM+L3 configuration
with half of its baseline DRAM bandwidth (0.5$\times$RD+0.5$\times$WR
which sums to 2.7TB/s) up through infinite bandwidth.
Figure~\ref{fig:l3_bw} summarizes the geometric mean
speedup of the MLPerf applications (training and inference),
normalized to the baseline GPU-N (without L3)\@. The performance
increases substantially when the L3 bandwidth grows from
0.5$\times$RD+0.5$\times$WR to 2$\times$RD+2$\times$WR, with
diminishing returns beyond that point. Overall, the UHB bandwidth
configurations of 2$\times$RD+2$\times$WR (total 10.8TB/s) comes
within 3\% and 6\% of the unlimited bandwidth settings for training and
inference respectively. This bandwidth is
well within the capabilities of next generation 2.5/3D interconnect technologies
assumed in our designs.

\begin{figure*}[t]
  \begin{subfigure}[t]{1\textwidth}
    \center
    \includegraphics[width=1\linewidth]{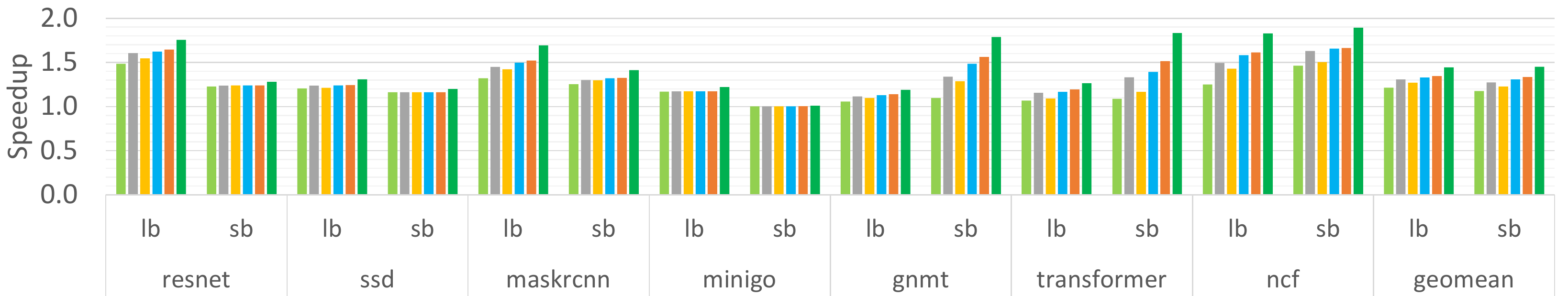}
    \caption{Training}
    \label{fig:perf_arch_train}
  \end{subfigure}
  \begin{subfigure}[t]{1\textwidth}
    \center
    \includegraphics[width=1\linewidth]{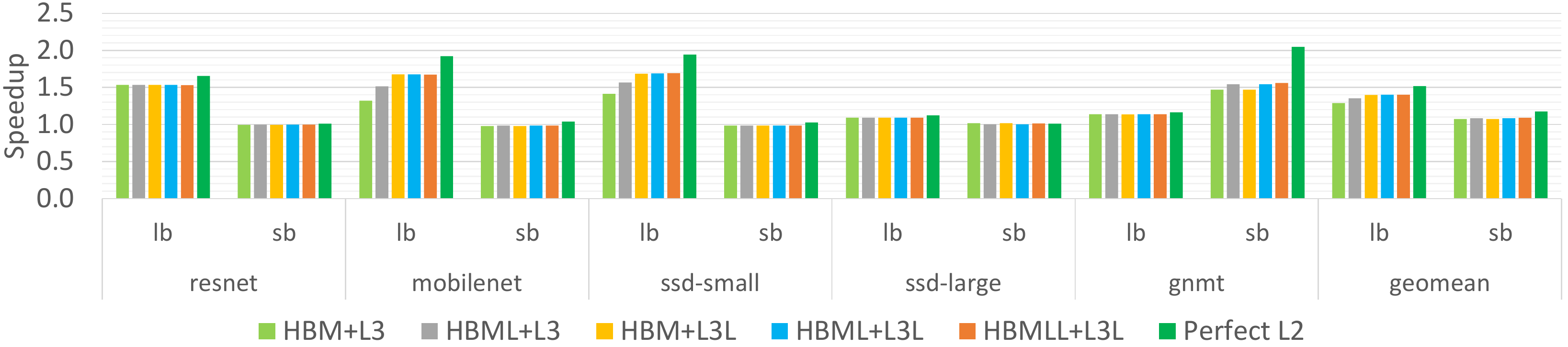}
    \caption{Inference}
    \label{fig:perf_arch_inference}
  \end{subfigure}
\caption{Performance of DL-optimized COPA-GPUs
with varying designs, normalized to the
baseline GPU-N performance. Large-batch and small-batch cases are
denoted by \textit{lb} and \textit{sb}.}
\label{fig:perf_arch}
\vspace{-0.2in}
\end{figure*}

We also evaluate the effect of the UHB link latency by varying the
total round-trip latency between the L2 and L3 caches from
0.25$\times$ to 1$\times$ of the DRAM access latency. Our experiments
(not shown) indicate that COPA-GPU architecture is not sensitive to L3
latency, as performance changes by less than 2\% across this latency
range. As a result, we set the UHB link bandwidth to be
2$\times$RD+2$\times$WR, for a total of 10.8TB/s and the round-trip
latency between L2 and L3 (UHB link latency plus L3 access latency) to
be half of the DRAM access latency for all the COPA-GPU configurations
with L3 that are listed in Table~\ref{tab:arch}.

Figure~\ref{fig:perf_arch} summarizes the MLPerf training and
inference performance of the COPA-GPU designs from
Table~\ref{tab:arch} in large batch (\textit{lb}) and small
batch (\textit{sb}) settings. Unsurprisingly, larger
cache capacity and higher memory bandwidth universally improve both large-batch and small-batch
training performance. For example, the additional cache capacity in
HBM+L3 improves large-batch and small-batch performance by 21\% and
18\% respectively, making it an attractive 3D COPA-GPU design.
Further, the additional HBM bandwidth provided by the HBML+L3 configuration
achieves an overall 31\% and 27\% speedup, making it an attractive
2.5D COPA-GPU design choice. HBM+L3L doubles L3 capacity instead of
scaling DRAM bandwidth and results in lower speedups than HBML+L3 (by
4\%) at both large- and small-batch scenarios, indicating that
increasing L3 capacity alone is not the best solution to scaling
training. Finally, ambitious DRAM options such as HBML+L3L and
HBMLL+L3L result in marginal performance gains (4\%) when
compared to HBML+L3 and do not justify its additional cost. 

For large-batch inference, the large L3 capacity in HBM+L3 improves
the performance by 29\% and HBM+L3L can reach speedup as high as
40\%. With 1,920MB of L3, further increases in HBM bandwidth is not
beneficial because most of the DRAM traffic has already been filtered out
by the large L3 cache. In fact, though not shown, we
find that DRAM bandwidth could be even reduced by 50\% without
affecting the performance. For small-batch inference, because the
performance saturation point is at 240MB of LLC, the performance
improvement via the combination of L3 and HBM is just 9\%.

Overall, we conclude that HBML+L3, which combines a substantially larger L3
cache (960MB) and moderately higher DRAM bandwidth (4.5 TB/s), is
likely the most optimal COPA-GPU design that will perform well for both DL
training (with 31\% gain at large-batch and 27\% gain at small-batch) and 
inference (with 35\% gain at large-batch and 8\% gain at small-batch)
without adding significant unnecessary cost through overprovisioning of memory
resources.  Inference is more sensitive to cache capacity and less sensitive to
DRAM bandwidth than DL training, and trading off additional HBM
resources for increased LLC capacity could be a plausible strategy if designing
a COPA-GPU specialized for just DL inference.

A large COPA-GPU enabled L3 reduces the total
DRAM-related per-GPU energy consumption by up to 3.4$\times$, as shown
in Section~\ref{sec:sec:uarch}\@. However, the improved DL-optimized
COPA-GPU utilization may lead to increased total design power that may not
be entirely mitigated by the power reduction within the memory
system. To mitigate growing thermal density, we expect future
high-end GPU systems will rely on liquid cooling technologies to enable
increased thermal envelopes compared to those possible
today~\cite{GoogleCooling,LiquidCooledSpec}.

\begin{figure}[t]
  \center
  \includegraphics[width=1\linewidth, trim= 0 0 0 0,clip]{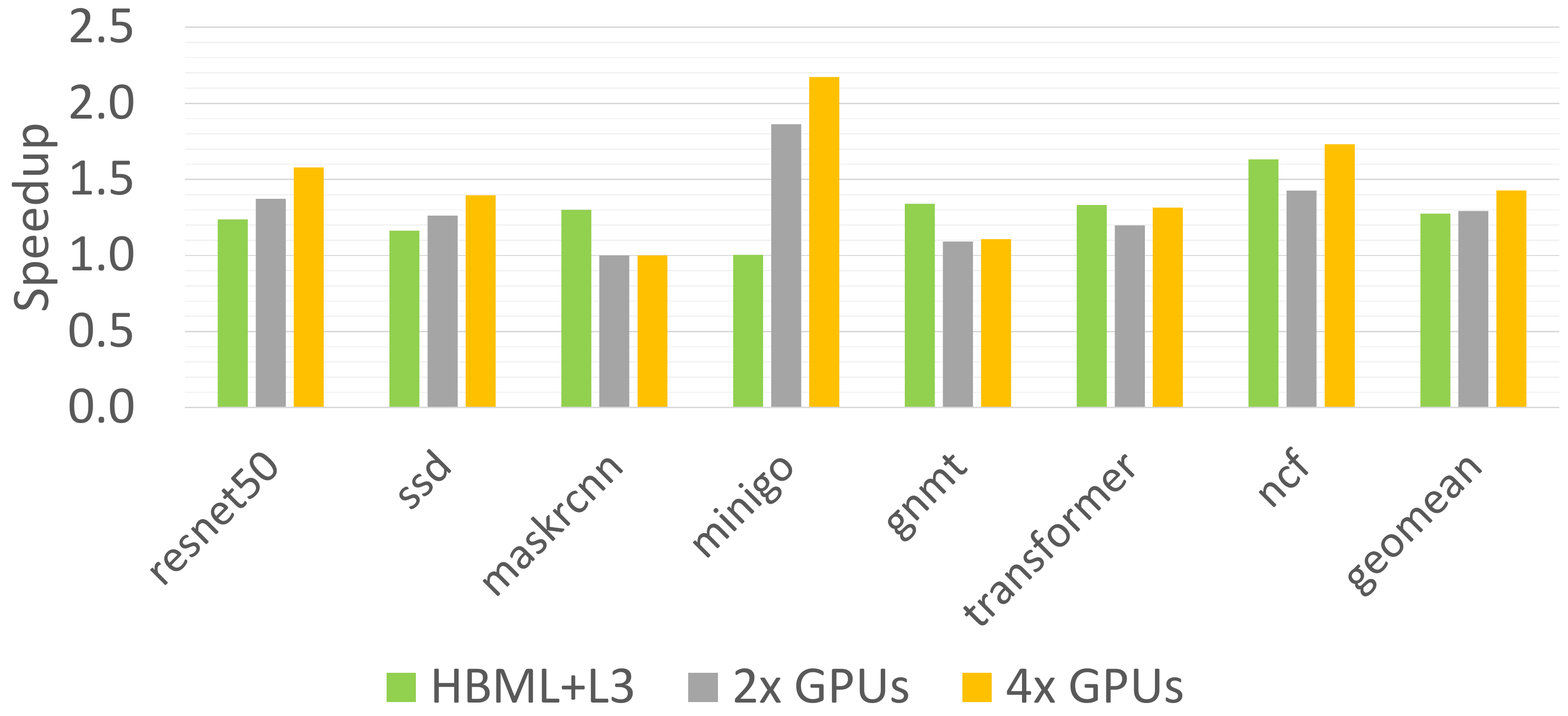}
  \caption{Performance of DL-optimized HBML+L3 COPA-GPUs versus additional data parallelism via 2$\times$GPU-Ns and 4$\times$GPU-Ns, normalized to the baseline GPU-N performance.}
  \label{fig:2dp}
  \vspace{-0.1in}
\end{figure}

\subsection{COPA-GPU Training Cost Efficiency at Scale}
\label{sec:sec:scale_out}

At scale, multi-GPU training performance is eventually limited by decreasing
per-GPU compute efficiency and growing system-level synchronization
overheads. To understand the scale-out efficiency of DL-optimized
COPA-GPUs, we compare the performance of an HBML+L3 COPA-GPU based system with
2$\times$ and 4$\times$ larger systems comprising the baseline
GPU-Ns. We fix the global batch size across all
configurations to maintain the same statistical
efficiency~\cite{hp_micro_2019}\@. Consequently, the per-GPU batch
size drops to one half and one quarter for the baseline GPU-N
configurations. We do not extrapolate the additional overheads
of distributed gradient synchronization at larger scales, and thus our 
analysis favors the 2$\times$ and 4$\times$ GPU-N
configurations.

Figure~\ref{fig:2dp} shows that doubling and quadrupling the number of
baseline GPU-N instances (2$\times$ GPU-Ns and 4$\times$ GPU-Ns)
results in mean 29\% and 43\% performance gains respectively for our
training workloads.  We find that a DL-optimized
HBML+L3 COPA-GPU configuration (with 27\% performance gain) provides similar levels of performance
to 2$\times$ GPU-Ns, yet should cost significantly less than buying and
hosting 2$\times$ larger installations of traditional GPU-Ns.
When compared to GPU-N, even though HBML+L3 doubles the area by adding
close to 826mm$^2$ of silicon, it is dominated by regular SRAM cache
arrays that are expected to achieve high manufacturing yields due to built-in
redundancy and error recovery. Moreover, this aggregate area is split into two smaller dies,
resulting in substantially lower cost per mm$^2$\@. HBML+L3 integrates 1.6$\times$ more HBM
memory, resulting in total aggregate cost lower than 2$\times$ of
GPU-N\@.  Thus, DL-optimized COPA-GPUs will provide
substantially better cost-performance at scale, saving on not just
overall GPU cost but additional system-level collateral such as
datacenter floorspace, CPUs, network switches, and other peripheral
devices.

\vspace{-0.05in}
\section{Related Work}
\label{sec:related}
Leveraging high capacity and high bandwidth on-chip and on-wafer
caches, or scratch memories, to store DL weights and activations has
been well explored for DL training/inference architectures.
DaDianNao~\cite{DaDianNao_micro_2014} was designed with 36MB of
on-chip eDRAM to cache model weights. Google's
TPUv1~\cite{TPU_isca_2017} allocated 28MB of on-chip memory mostly for
caching activations, while TPUv2 and TPUv3~\cite{TPU_2020} increased
it to 37MB\@.  Recent DL accelerators such as Graphcore's
IPU~\cite{graphcore_2020}, Groq's TSP~\cite{groq_2020}, Alibaba's
HanGuang~\cite{hanguang_isscc_2020}, and Cerebras's
WSE~\cite{cerebras_2019} replaced off-chip memory with hundreds of MBs
of high-bandwidth on-chip SRAM to satisfy increasing memory bandwidth
requirements. Unlike domain specific accelerators that
are highly tuned for DL workloads, COPA-GPUs provide high levels of
GPU design reuse across application domains, while also enabling
memory-system specialization for individual domains.

Multi-chip module (MCM) packaging has been extensively studied and
deployed to integrate heterogeneous and homogeneous chips within a
package, aiding the scaling of compute and memory bandwidths for a
wide variety of legacy GPU and CPU applications. Prior work~\cite{MCM-GPU,NUMA-GPU, EnergyMCM} 
has focused on developing MCM-GPU
architectures to strong scale GPU performance beyond the limitations
of a single monolithic die by leveraging on-package and on-board
integration technologies. In a follow-on work~\cite{HMG_hpca_2020,Young18}, the
authors extend MCM-GPU architectures with advanced caching and HW/SW
cache-coherency protocols to overcome NUMA
limitations. MCM-3D-NoC~\cite{3DNOCS} tackled the interconnect
scalability issues of MCM integration over active interposers. In the
CPU space, recent AMD CPU architectures~\cite{AMDZEPPELIN,
AMD-chiplets-isscc20} leverage multi-module on-board integration to
provide scalable and modular CPU architectures. Finally, Kannan et
al.~\cite{Kannan15} proposed to disaggregate large monolithic CPU
designs into smaller chips for cost
reduction. In~\cite{simba_micro_2019}, the authors propose and
quantify the costs and benefits of using MCMs with fine-grained
domain specific chiplets for DL inference.

General purpose CPU SoCs have already leveraged MCM designs with very large 
on-package eDRAM caches. Intel improved its mobile CPU performance by
combining a CPU SoC and 128MB of on-package eDRAM organized as a
victim cache~\cite{IrisPro}\@ and IBM scaled up its Gen-Z mainframe
performance with 960MB of L4 eDRAM cache~\cite{EDRAM-IBM-ISSCC20}\@.
While a COPA-GPU leverages the previously proposed concepts of on-package
integration and large caches, our work is (1) the first to identify
and solve the diverging architectural requirements between FP32 (or
larger) based HPC and FP16 (or smaller) based DL workloads in GPUs
and (2) the first to develop the reusable  GPU architecture concept,
enabling cost-effective  GPU domain-specialization for HPC and deep learning.

\vspace{-0.05in}
\section{Conclusion}
\label{sec:conclusion}

In this work, we demonstrate that diverging architectural requirements between the
HPC and DL application domains put converged GPU designs on a
trajectory to become significantly under-provisioned for DL and
over-provisioned for HPC. We propose a new composable GPU architecture
that leverages emerging circuit and packaging technologies to provide
specialization, while maintaining substantial compatibility across
product lines. We demonstrate that COPA-GPU architectures can enable selective deployment of
on-package cache and off-chip DRAM resources, allowing manufacturers
to easily tailor designs to individual domains.
Our analysis shows that DL-optimized COPA-GPUs will provide 
impressive per-GPU training and inference performance
improvements, while still efficiently supporting scaled-down
HPC-targeted designs. DL-optimized COPA-GPUs will also result in
reduced cost datacenter cost by minimizing the number of GPUs
required to achieve scale-out
training performance targets, making COPA-GPU an attractive paradigm for 
increasing individual and aggregate GPU
performance without over-optimizing the product for any specific domain.



\bibliographystyle{IEEEtranS}
\bibliography{refs}

\end{document}